\documentclass[a4paper,fleqn,usenatbib]{mnras}

\usepackage[T1]{fontenc}
\usepackage{ae,aecompl}

\usepackage{graphicx}	% Including figure files
\usepackage{amsmath}	% Advanced maths commands
\usepackage{amssymb}	% Extra maths symbols

\newcommand{\disp}{\displaystyle}

\usepackage[usenames,dvipsnames]{xcolor}
\title[Sound speed and Titan's lakes composition]{How speed of sound measurements could bring constraints on the composition of Titan's seas}

\author[D. Cordier]{
D.~Cordier,$^{1}$\thanks{E-mail: daniel.cordier@univ-reims.fr}\\
$^{1}$Groupe de Spectrom\'{e}trie Mol\'{e}culaire et Atmosph\'{e}rique,\\
UMR 6089, Campus Moulin de la Housse, BP 1039,\\
Universit\'{e} de Reims Champagne-Ardenne,\\
51687 Reims, France\\
}
\date{Accepted XXX. Received YYY; in original form ZZZ}

\pubyear{2016}

\begin{document}
\label{firstpage}
\pagerange{\pageref{firstpage}--\pageref{lastpage}}
\maketitle

\begin{abstract}
  {The hydrocarbons seas of Titan, discovered by Cassini/Huygens mission are among the most mysterious and interesting features
  of this moon. In the future, a possible dedicated planetary probe, will certainly measure the speed of sound in this 
  cryogenic liquid, as it was planned in the case of a Huygens landing into a sea.}
  {Previous theoretical studies of such acoustic measurements were based on rather simple models, leading in some cases to unphysical
  situations. Employed in a vast body of chemical engineering works, the state of the art PC-SAFT model has been recently introduced in studies
  aimed at Titan. Here, I revisit the issue of the speed of sound in Titan's liquids, in the light of this theory.
  I describe, in detail, the derivation of the speed of sound from the chosen equation of state and the potential limitations
  of the approach. To make estimations of the composition of a ternary liquid mixture N$_2$:CH$_4$:C$_2$H$_6$ from speed of sound measurements
  an original inversion algorithm is proposed. It is shown that $50$ measures between $90$ K and $100$ K are enough to ensure an accuracy of the derived
  compositions better than $10$\%. The influence of the possible presence of propane is also investigated.}
\end{abstract}


%%%%%%%%%%%%%%%%%%%%%%%%%%%%%%%%%%%%%%%%%%%%%%%%%%%%%%%%%%%%%%%%%%%%%%%%%%%%%%%%%%%%%%%%%%%%%%%%%%%%%%%%%%%%%%%%%%%%%%%%%%%%%%%%%%%%
\section{Introduction}
%\begin{linenumbers}
%\modulolinenumbers[5]

%\linenumbers
	Among a multitude of fascinating features, Titan, the main satellite of Saturn, is -- with the Earth -- the only body of the solar
system bearing stable liquid phases at its surface.
    These hydrocarbon seas and lakes remain largely mysterious: they appear amazingly flat \citep{zebker_etal_2014},
with unexplained reflectivity events \citep{hofgartner_etal_2014}, while their precise chemical composition is not well 
known. Only the presence of ethane has been detected \citep[][]{brown_etal_2008} and estimations performed with numerical models 
\citep[][]{tan_etal_2013,glein_shock_2013,cordier_etal_2009,cordier_etal_2013a} somewhat disagree.
    Given their important role as reservoir in the hydrocarbons cycle, and because of their high exobiological potential 
\citep[][]{mckay_smith_2005,schulzemakuch_grinspoon_2005,lunine_2010}, these lakes/seas would be very interesting targets for an
{\it in situ} exploration. Already mission concepts are studied: for instance the project 
TiME \citep[{\it Titan Mare Explorer}, see][]{stofan_etal_2011} proposes a sea surface exploration with a boat while \cite{lorenz_etal_2015} suggest
a submarine. Both proposals incorporate an instrument that exploites the properties of sound propagation in liquids.

  The idea of instruments based on acoustic measurements has been investigated and implemented.
% \PL{For} instance \cite{hanel_strange_1966} \PL{studied} the possibility of CO$_2$, N$_2$ and Ar abundances determination in the atmosphere
of Mars. Originally aimed for thunder detection, sound sensors have been deployed at the surface of Venus  
\cite[][]{ksanfomality_etal_1986a,ksanfomality_etal_1986b}.
Finally, \cite{lorenz_1999} has discussed what could be learned about outer planet
atmospheres using acoustic properties. 
  In the context of Titan, the {\it Huygens} probe was equipped with the instrument called {\it Acoustic Properties Investigation} (API), belonging
to the {\it Surface Science Package}, which 
consisted of two units:  API-V (Velocity of sound) and API-S (Sounding) \citep[][]{svedhem_etal_2004}. 
Even if this instrument had also capabilities for performing analysis in the atmosphere, it was mainly aimed at liquid phase
investigations. Since {\it Huygens} landed on a dry region, no measurements were done in Titan's surface liquid. However acoustic
data collected during the descent allowed \cite{hagermann_etal_2007} to derive constraints on the atmosphere composition.\\

   \cite{hagermann_etal_2005} did a conceptual work of how the measurements, acquired by
\textit{Huygens}' instruments in a cryogenic liquid, could given information regarding its composition. Among other physical quantities,
they considered the speed of sound. Facing the lack of published speed of sound measurements, particularly for the hydrocarbons mixtures, 
they used an equation of fit.
They validated the latter by comparison with the results given by the equation of state (hereafter EoS) published by \cite{peng_robinson_1976}. 
They also employed data coming from the National Institut of Standards and Technology (NIST14). However, they did not give the full 
derivation of the computed speed of sound.
In the context of depths sounding, \cite{arvelo_lorenz_2013} made computations based on the speed of sound in cryogenic liquids. Their
work also involves the NIST14 database. Unfortunately, the obtained velocities exhibit discontinuities which are certainly not physical \citep[see figure 3 of][]{arvelo_lorenz_2013}.
In addition, these authors assumed that their adopted linear fit can be extrapolated to temperatures higher than $92.5$ K.
For all these reasons, I decided to explore the question of speed of sound in cryogenic liquids and to revisit its sensitivity to
chemical composition. 
Successfully used in countless works of chemical engineering, and introduced in Titan research field by \cite{tan_etal_2013},
I have chosen to use the up-to-date, Helmholtz energy based theory, 
PC-SAFT\footnote{Perturbed-Chain Statistical Associating Fluid Theory}.
On average, PC-SAFT is more accurate than all other EoS, and particularly Peng-Robinson cubic EoS 
\citep[see for instance][]{diamantonis_etal_2013,annesini_etal_2014}. Focusing on the speed of sound dependency with
chemical composition, I describe a concept of a very simple instrument, similar to the API-V, but including an active
temperature control of the probed liquid. In this paper, I emphasize the required accuracy and on the needed number of velocity of sound measurements; this,
in a way as quantitative as possible. In addition, the exact derivation of the speed of sound from PC-SAFT quantities is presented in detail.
The system is not thought to compete with an instrument as accurate as a mass spectrometer, 
but it would be very useful in the case of a failure of such an high precision sensors.
  In Sect.~\ref{measur} I discuss the principle of the speed of sound measurements. Sect.~\ref{calspeed}
is devoted to the thermodynamical computations and model description. In Sect.~\ref{datainv}, the inversion algorithm is described and the sensibility of
results to experimental conditions is discussed. Final remarks and conclusion will be made in Sect.~\ref{concl}.
  
%%%%%%%%%%%%%%%%%%%%%%%%%%%%%%%%%%%%%%%%%%%%%%%%%%%%%%%%%%%%%%%%%%%%%%%%%%%%%%%%%%%%%%%%%%%%%%%%%%%%%%%%%%%%%%%%%%%%%%%%%%%%%%%%%%%%
\section{The principle of the measurements}
\label{measur}

\begin{figure}
\begin{center}
\includegraphics[width=8 cm]{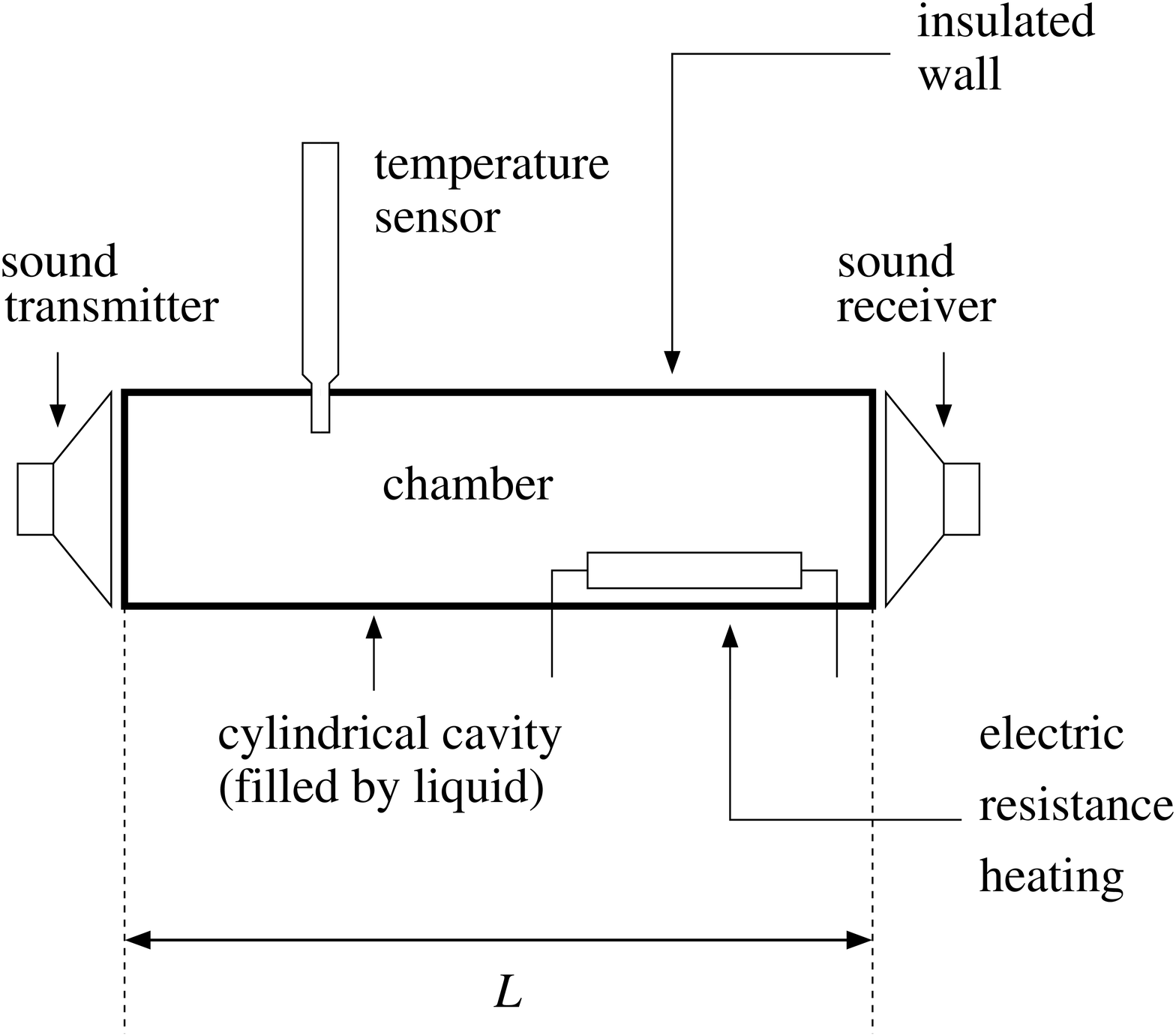}
\caption[]{\label{kundt_tube}Sketch of the device that could be used to measure the speed of sound in the liquid of a Titan's 
         lake. The temperature control of the liquid is ensured by means of temperature sensors and an electric heating system.}
\end{center}
\end{figure}

  If the bulk composition of Titan's lakes contains only three compounds, characterizing the composition can be achieved 
if we know the mole fraction of nitrogen $x_{\rm N_{2}}$  and the ratio $r_{46}= x_{\rm CH_4}/x_{\rm C_2H_6}$, where 
$x_{\rm CH_4}$ and $x_{\rm C_2H_6}$ are respectively the mole fractions of methane and ethane. As shown, in principle, by \cite{hagermann_etal_2005},
an appropriate set of physical property measurements can be used as an indicator of the chemical composition of the considered
liquids. Following this idea, the measurements of two independent values of the speed of sound $u$ in a liquid could
be used -- at least formally -- to determine the values of $x_{\rm N_{2}}$ and $r_{46}$. These
measurements could be performed for two different temperatures, and more extended set of measurements will bring stronger
constraints on the composition. If $u_{1}$ and $u_{2}$ are respectively the determined sound velocity at 
temperature $T_{1}$ and $T_{2}$, one has to solve the set of equations (\ref{system1}).  
\begin{equation}\label{system1}
\begin{array}{lcl}
  u(x_{\rm N_{2}}, x_{\rm CH_{4}}, x_{\rm C_{2}H_{6}}, T_{1}) &=& u_{1} \\
  u(x_{\rm N_{2}}, x_{\rm CH_{4}}, x_{\rm C_{2}H_{6}}, T_{2}) &=& u_{2} \\
\end{array}
\end{equation}
  This assumes that the liquid mixture does not evaporate at a temperature in the range $[T_{1},T_{2}]$. Moreover the liquid has to be 
isolated from any vapor phase such as the local atmosphere.
  With more than two measurements, systems similar to (\ref{system1}), each corresponding to a couple $(T_{i},T_{j})_{i\ne j}$, can
be resolved. The composition is then obtained by averaging the inferred mole fractions.
  In the frame of this approach, a law $u(x_{i},T)$ has to be determined during the preparation
of the space mission, by theoretical works and/or laboratory calibrations.
  One might conceive of a run of measurements in a passive mode, {\it i.e.} taking advantage of the natural variations of the surface
temperature of the sea. Unfortunately, the expected amplitude of climatic temperature variations is of the order of a few kelvins
by Titan's year ($29.5$ Earth years); these circumstances make these variations unusable for our purpose. Thus, an active system of
temperature control is needed.
  A device inspired by the Kundt's tube \citep{kundt_1866} could be employed. Althought it would be equipped by valves, such a system has the great 
advantage of a small number of mobile mechanical pieces. The Kundt's device is sketched in Fig.~\ref{kundt_tube}, 
%
%===================================================================================================================================
%===================================================================================================================================
%
enclosing the tube (that could also be called ``chamber of measurement'') an insulating wall prevents too high a heat flux
towards ambient liquid, and stabilizes the temperature when the acoustic experiments are conducted. The probed liquid is heated
by an electrical resistor whereas a device and/or a dedicated procedure ensure the homogeneity of the temperature in
the chamber. The measurement can be done by determining the time taken by an ultrasonic pulse to travel the length of the tube
between the emitter and the receiver. This is the operating principle achieved by API-V {\it Huygens} or ``SOSO'' ({\it TiME}).
I emphasize that the tube could also be used in a way similar to what Kundt has done. Indeed, the tube is a cylindrical resonant cavity
(of length $L$)
 The natural frequencies $f_{n}$ of such a cavity obey the simple law given by Eq. (\ref{fn})
\begin{equation}\label{fn}
  f_{n}= n \, \frac{\disp u}{\disp 2 L}
\end{equation}
\citep[see][]{feynman_etal_1963}. For a fixed cavity length $L$, the determination of only two consecutive resonant 
frequencies gives the speed of sound: $u= 2L (f_{n+1}-f_{n})$.
The measurement of a travel time is then replaced by a detection of resonance frequencies.
The larger the sample of acquired data is, the more precise the value of the speed $u$ is. An electronic unit dedicated
to frequency generation and a signal receiver processing are required to complete the system. In addition, the liquid heating 
could be easily done with the help of an electrical resistor, while temperature measurements would be monitored by means of thermocouples. The
global performances ({\it e.g.} number of measurements per unit of time) determine the actual accuracy achieved for the law
$u(T)$.\\
 In the next section, we discuss the precision needed for speed of sound determinations (or equivalently for frequencies) to get results
useful for liquid composition.
   Concerning the range of temperature that would be used, I propose the interval from 90 K to 100 K. Indeed, the ground
temperature of Titan's polar regions has been estimated to be $\sim 90$ K based on near-surface brightness temperature 
measurements \citep[][]{jennings_etal_2009}. In addition, methane has its boiling point at $111.2$ K. It is surely technically 
easier to heat the liquid of lake up to 100 K than to cool it down to $80$ K, temperature at which the fluid might solidify.

%%%%%%%%%%%%%%%%%%%%%%%%%%%%%%%%%%%%%%%%%%%%%%%%%%%%%%%%%%%%%%%%%%%%%%%%%%%%%%%%%%%%%%%%%%%%%%%%%%%%%%%%%%%%%%%%%%%%%%%%%%%%%%%%%%%%
\section{Estimation of the speed of sound in a ternary mixture of nitrogen, methane and ethane }
\label{calspeed}

The speed of sound in a given substance is computed from its thermodynamic properties using the general equation
\citep[see for instance][]{diamantonis_economou_2011}
   
\begin{equation}\label{speed1}
   u = \sqrt{\frac{C_{\rm P}}{C_{\rm V}} \left(\frac{\partial P}{\partial \rho}\right)_{T}}
\end{equation}

  with $C_{\rm P}$ and $C_{\rm V}$ respectively the isobaric and isochoric specific heat, $P$ is the pressure, $\rho$
the density and $T$ the temperature.
  This equation can be easily derived from first principles and classical thermodynamics equations 
\citep[{\it e.g}.][]{lobo_ferreira_2006}. 
      The Helmholtz free energy is generally used in statistical thermodynamics to express EoS, since most properties of
interest can be obtained by proper differentiation of it. With PC-SAFT, the total Helmholtz energy -- denoted $A$ -- can be written as a sum 
\begin{equation}\label{A}
   A= A^{(\rm id)} + A^{(\rm res)}
\end{equation}
%  
%===================================================================================================================================
\begin{figure}
\begin{center}
\includegraphics[width=6 cm, angle= -90]{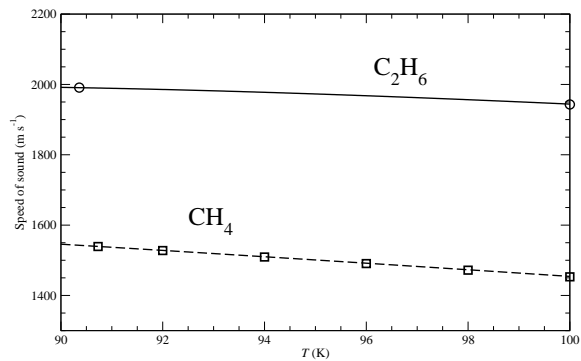}
\caption[]{\label{compamodexpe}Comparison between experimental determination of the speed of sound (symbols) for 
           CH$_4$ and C$_2$H$_6$ and the outputs of our model. Squares: experimental data from \cite{setzmann_wagner_1991},
           circles: data taken in \cite{younglove_ely_1987}.}
\end{center}
\end{figure}
%===================================================================================================================================
%
  in which $A^{(\rm id)}$ represents the Helmholtz energy of the corresponding ideal-gas,
$A^{(\rm res)}$ being the residual contribution that accounts for intermolecular interactions.
In the PC-SAFT framework molecules are conceived to be chains comprised of freely jointed spherical segments. Inter-segment
and intermolecular potentials are introduced and the theory provides the resulting Helmholtz energy of the macroscopic system.
 The reader who is 
interested in the nature of the terms included in $A^{(\rm res)}$ is invited to consult the vast literature devoted to the basis
of PC-SAFT. The foundations of this theory have been originally published by \cite{gross_sadowski_2001}. A good introduction
to PC-SAFT can be found, in \cite{soo_2011}.
This theory has been extensively tested in the context of cryogenic liquids by \cite{tan_etal_2013} and
it has also been proved that it reproduced laboratory data, particularly isotherms and binary diagrams, with a very satisfying degree of accuracy. 
However, despite its great performances, PC-SAFT, similarly to the theories applied by \cite{cordier_etal_2009} or 
\cite{glein_shock_2013}, include free parameters that have to be adjusted; but theories belonging to the ``SAFT family'' have the 
advantage to rely on a strong statistical physics basis.
Moreover, I have checked that PC-SAFT 
reproduces the reference ternary mixture (N$_2$, CH$_4$, C$_2$H$_6$) of \cite{gabis_1991} provided as supplementary data by \cite{glein_shock_2013}.\\
 In the
frame of this EoS, each considered species is characterized by the PC-SAFT three parameters, namely the segment diameter $\sigma$,
the depth of the potential $\epsilon/k_{B}$, and the number of segments per chain $m$; the values of those parameters have all been
given in \cite{tan_etal_2013,cordier_etal_2016b}. Beside this, the binary interaction parameters $k_{ij}$ 's are also taken in this article.
  Eq. (\ref{speed}) can be rewritten as
\begin{equation}\label{speed}
   u = \sqrt{\frac{C_{\rm P}}{C_{\rm V}} \frac{1}{k_{T} \rho}}
\end{equation}
where $k_{T}^{-1} = \rho (\partial P/\partial\rho)_{T}$. The density $\rho$ and 
the derivative $(\partial P/\partial\rho)_{T}$ are provided by PC-SAFT, 
whereas the non-ideal isobaric heat capacity of the mixture $C_{\rm P}$ is derived using the relation
%
%===================================================================================================================================
\begin{figure}
\begin{center}
\includegraphics[width=8 cm, angle= 0]{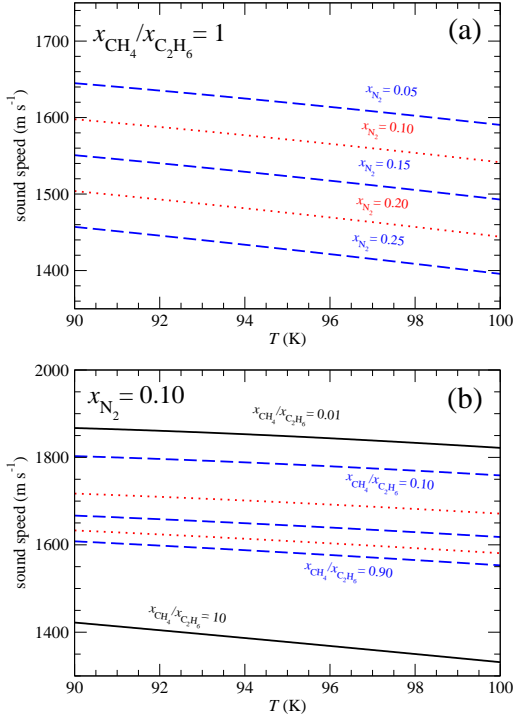}
\caption[]{\label{figCsound}The calculated speed of sound as a function of temperature in the considered ternary mixture: (a) the
           ratio $r_{46}= x_{\rm CH_{4}}/x_{\rm C_{2}H_{6}}$ is fixed at $1$ while the nitrogen mole fraction $x_{\rm N_{2}}$ takes the value
           $0.05$, $0.10$, $0.15$, $0.20$ and $0.25$. (b) The mole fraction of nitrogen is set to $0.10$ while the ratio
           $r_{46}= x_{\rm CH_{4}}/x_{\rm C_{2}H_{6}}$ takes the values $0.01$, $0.10$, $0.30$, $0.50$, $0.70$, $0.90$ and $10$. 
           For all calculations, the pressure has been fixed at $1.5$ bar, the observed value on Titan's surface
           \citep[][]{niemann_etal_2005}.}
\end{center}
\end{figure}
%===================================================================================================================================
%
\begin{equation}\label{CpmCv}
   C_{\rm P} - C_{\rm V} = \frac{\disp T \, \alpha^{2}}{\disp k_{T} \, \rho} \, \bar{M}
\end{equation}
where $\alpha = k_{T} (\partial P/\partial T)_{V}$ is also computed with PC-SAFT. $\bar{M}$ represents the average molar mass of the mixture.
The isochoric heat capacity $C_{\rm V}$ is given by \citep[see][Eq. 19]{diamantonis_economou_2011}
\begin{equation}\label{Cv}
   C_{\rm V} = \underbrace{-T \, \left(\frac{\disp\partial^{2} A^{(\rm id)}}{\disp\partial T^{2}}\right)_{\rm V}}_{= C_{V}^{(\rm id)}}
               -T \, \left(\frac{\disp\partial^{2} A^{(\rm res)}}{\disp\partial T^{2}}\right)_{\rm V}
\end{equation}
The term denoted $C_{V}^{(\rm id)}$ is the isochoric heat capacity of the corresponding ideal gas.
 I made a series of tests to evaluate this term. Among them, I used the group-contribution method 
developed by Joback and Reid \cite[][]{joback_1984,joback_reid_1987} and summarized in \cite{poling_2007}. 
This approach consists of an approximate estimation of the requested thermodynamical quantities. Unfortunately, doing so did not
yield to speeds of sound in very good agreement with the tabulated experimental data.
 As a consequence, I found that adjusting the individual heat capacities $C_{P, i}^{(\rm id)}$ by fitting the individual 
velocity of sound data, gives much better results. 
This isochoric $C_{V,i}^{(\rm id)}$ is derived from $C_{P,i}^{(\rm id)}$ thanks to Mayer's law; while $C_{V}$ comes from Eq. (\ref{Cv}),
Eq. (\ref{CpmCv}) provides $C_P$.
A comparison between the model
outputs and experimental data is presented in Fig.~\ref{compamodexpe}. The agreement for methane speed
of sound \citep{setzmann_wagner_1991}; and also for ethane data \citep{younglove_ely_1987} appears very good.
I have also been able to reproduce the velocity of sound in liquid nitrogen at $77$ K \cite[859 m s$^{-1}$, ][]{zuckermawar_mazel_1985}. 
In Fig.~\ref{figCsound} the calculated
speed of sound for temperatures between $90$ K and $100$ K, is plotted for ranges of mole fractions that cover the plausible abundances of Titan's seas. 
As it can be seen in panel (a) and (b) of this figure, $u$ always
decreases when $T$ increases. It is noticeable that the amplitude of these variations, between the boundaries of the considered
interval, is roughly $\sim 50-60$ m s$^{-1}$. 
%
%===================================================================================================================================
\begin{figure}
\begin{center}
\includegraphics[width=6 cm, angle= -90]{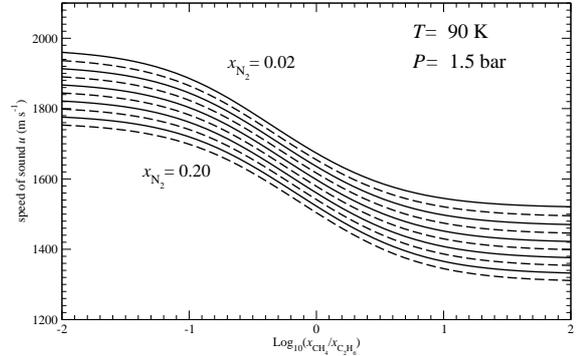}
\caption[]{\label{speedVSratio90K}Tabulated speed of sound versus ratio $x_{\rm CH_{4}}/x_{\rm C_2H_6}$ at $T= 90$ K and 
           $P= 1.5$ bar, for mole fraction of nitrogen $x_{\rm N_2}=$ $2 \times 10^{-2}$, $4 \times 10^{-2}$, $6 \times 10^{-2}$, 
           $8 \times 10^{-2}$, $1 \times 10^{-1}$, $1.2 \times 10^{-1}$, $1.4 \times 10^{-1}$, $1.6 \times 10^{-1}$, 
           $1.8 \times 10^{-1}$ and $2 \times 10^{-1}$.}
\end{center}
\end{figure}
%===================================================================================================================================
%
This implies an accuracy of measurements of a few percents, if we want to capture the variations of $u$ with temperature.
The curves in Fig.~(\ref{figCsound}) do not exhibit local extrema, this is 
a useful (and expected) feature in the perspective of data inversion. Finally, we can remark that, not surprisingly, the computed
speed of sound gets higher when the average molar mass increases.

%%%%%%%%%%%%%%%%%%%%%%%%%%%%%%%%%%%%%%%%%%%%%%%%%%%%%%%%%%%%%%%%%%%%%%%%%%%%%%%%%%%%%%%%%%%%%%%%%%%%%%%%%%%%%%%%%%%%%%%%%%%%%%%%%%%%
\section{Data inversion}
\label{datainv}

   One major goal of this discussion is to assess the possibility of deriving chemical composition information from acoustic
velocity. Thus, I have simulated a lake chemical composition extraction from artificial acoustic measurements. The samples
were constructed using a finite number of sound velocities, computed with the model, for a list of $N$ temperatures
in the range of interest, {\it i.e}. $90$--$100$ K.
In general, an inverse problem, like the one we are facing here, can be treated in different ways \citep[][]{aster_etal_2012}. 
In a first attempt, I have tried to minimize a likelihood function based on a $\chi^{2}$. However, a very bad convergence was
observed. Instead, I built pre-computed tables of speeds of sound depending on three parameters: the temperature $T$, the nitrogen
mole fraction $x_{\rm N_2}$ and the ratio $r_{46}= x_{\rm CH_{4}}/x_{\rm C_2H_6}$. The data inversion is performed using the
following algorithm:
  
\begin{enumerate}
  \item for each ``experimental'' temperature $T_{i}$, the value of the ratio $r_{46}$ is searched by solving the
        equation $u_{\rm exp}=u_{\rm Table}$, for each value of $x_{\rm N_2}$ implemented in the multidimensional table.
        In Fig.~\ref{speedVSratio90K}, I have displayed examples of curves giving $u$ versus $\mathrm{log}_{10}(r_{46})$ for a temperature
        fixed at 90 K and a nitrogen mole fraction ranging from $0.02$ to $0.20$.
        This way, one builds a set of curves providing $x_{\rm N_{2}}^{(T)}$ as a function of the ratio $r_{46}$.
%
%===================================================================================================================================
\begin{figure}
\begin{center}
\includegraphics[width=6 cm, angle= -90]{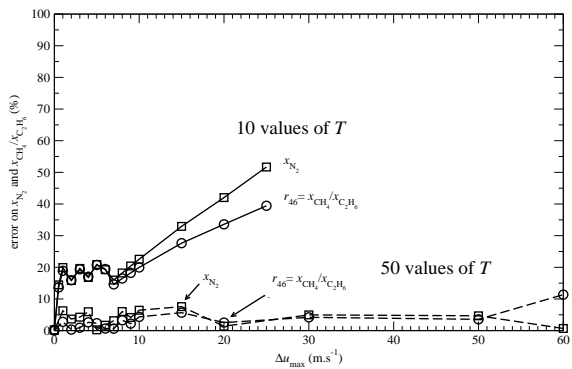}
\caption[]{\label{errorevol}Results of simulations of ``experimental'' data inversion. Two sets of ``experimental'' data have been analyzed: one 
           containing speeds of sound for $10$ values of temperature uniformly distributed between $90$ and $100$ K, a second consisting
           in 50 values. In order to mimic instrumental errors on speed of sound measurements, a random signal has been added to the data,
           originally computed using my PC-SAFT-based thermodynamic model. This signal consists of randomly chosen speed of sound contributions
           uniformly distibuted between $-\Delta u_{\rm max}$ and $+\Delta u_{\rm max}$, where $\Delta u_{\rm max}$ is the maximum measurement
           error allowed in the simulation. This plot represents the error on chemical composition determination as a function of $\Delta u_{\rm max}$.}
\end{center}
\end{figure}
%===================================================================================================================================
%
  \item in a second step, for each couple of temperature values $(T_{i}, T_{j})_{i\ne j}$, the coordinates of the 
        intersection point of the curves $x_{\rm N_2}^{(T_{i})}(r_{46})$ and
        $x_{\rm N_2}^{(T_{j})}(r_{46})$ are determined using a root finding method. It should be noted that, in an ideal situation, all the curves have to 
        intersect strictly at
        the same point $(\left.r_{46}\right|_{0}, \left.x_{\rm N_2}\right|_{0})$ characterizing the chemical
        composition of the studied liquid. In practice this is not the case, due to numerical errors, and above all
        because of experimental 
        uncertainties, the derived points are scattered around an average position. The barycenter of these points represents the
        observed chemical composition, while the scattering offers information on uncertainties associated with this derivation.      
\end{enumerate}
    Here, the uncertainties were estimated by computing the difference between the ``true'' composition used at the
time of the look-up-table construction, and the abundances inferred with our inversion algorithm.
    It is valuable to estimate the sensitivity of the results ({\it i.e}. derived values of the composition of the liquid) to the 
uncertainties on the measured speed of sound. In order to mimic the instrumental errors, I added a random signal to velocity values, originally 
computed with the thermodynamical model. Subsequently, the analysis of these surrogate samples is done by 
the use of the algorithm described above.
Thus, I applied artificial speed errors $\Delta u_{i}$ uniformly distributed between $-\Delta u_{\rm max}$ and $+\Delta u_{\rm max}$,
where $\Delta u_{\rm max}$ is an arbitrary chosen value, representing the maximum error allowed in the numerical test.
    As already noticed in Sect.~\ref{calspeed}, the accuracy of the speed of sound in the considered liquid must be much lower 
than $\sim 50-60$ m s$^{-1}$ for a typical value of speed around $1500$ m s$^{-1}$. Then the required instrumental precision should 
be a priori better than $\sim 50$ m s$^{-1}$.
   In Fig.~\ref{errorevol} we have reported the results of two simulations: one based on a set of $N=$10 values of temperature uniformly
distributed between 90 and 100 K, the second consisting in $N=$50 values. In this figure, the errors on nitrogen mole
fraction and errors on the ratio $r_{46}= x_{\rm CH_{4}}/x_{\rm C_2H_6}$ were plotted as functions of the value of the maximum error
$\Delta u_{\rm max}$. Both simulations were done using an initial ``true'' chemical composition corresponding to $x_{\rm N_2}= 0.10$
and $x_{\rm CH_{4}}/x_{\rm C_2H_6}= 0.50$.
As one can see, for $N= 10$ the errors are relatively stable around $\sim 20$\% for $\Delta u_{\rm max}$ up to $\sim 10$ m s$^{-1}$;
beyond this value the errors increase regularly and finally the algorithm no longer converges properly for $\Delta u_{\rm max} \sim 25$
m s$^{-1}$.
  This numerical experiment corresponds to a temperature control at a $1$-K level, which appears well feasible.
As a corollary, in that case, the speed of sound measurements have to be done with an absolute accuracy better than $10$ m s$^{-1}$,
which corresponds to a relative precision better than $10^{-2}$.
In Fig.~\ref{errorevol}, the $N= 50$ numerical experiment shows clearly that, for a given $\Delta u_{\rm max}$, the increasement
of the number of measurements yields to an appreciable improvement in the chemical composition determination. In addition, the results of
the inversion remains acceptable (errors remain below $\sim$10\%) up to $\Delta u_{\rm max}\sim 60$ m s$^{-1}$. Nonetheless, a number
of measurements of $N= 50$ requires a temperature control of the liquid at the level of $\sim 0.2$ K that is more difficult 
to achieve than a $\sim 1$ K level. This gives an idea of the precision needed on the temperature control.
The dependency on $N$ is anticipated since the final step of the inversion algorithm consists in an average. 
During the design stage of the instrument, a compromise will have to be found between temperature control and the accuracy of 
speed of sound determinations.\\
I have also checked that the errors on temperature have a negligible influence. For instance, in the
case of a sample of $10$ measurements between $90$ and $100$ K, randomly distributed errors on $T$ with a maximum of $\pm 0.2$ K lead to errors on speed of sound
below $0.8$ m.s$^{-1}$ while a maximum error of $\pm 0.5$ K produces speed uncertainties not larger than $2$ m.s$^{-1}$, value well below
the $\Delta u_{\rm max}$ values considered above. 
This demonstrates that, if the temperature accuracy required by a number of independent measurements can be guaranteed 
(\textit{e.g.} better than $1$ K for $10$ measurements between $90$ and $100$ K), then the effects of the inaccuracies in 
the temperature measurements on the results are negligible compared to the uncertainties in the speed of sound measurements.\\
    
    Finally, photochemical models of Titan's atmosphere show that propane could be produced \citep[][]{lavvas_etal_2008a,lavvas_etal_2008b}.
Thus, I introduced some amount of propane in the case where $\Delta u_{\rm max}$ is fixed at $30$ m s$^{-1}$. The chemical composition of
the ``experimental sample'', {\it e.g.} $x_{\rm CH_4}= 0.30$, $x_{\rm N_2}= 0.10$ and $x_{\rm C_2H_6}= 0.60$, is replaced by
 $x_{\rm CH_4}= 0.30$, $x_{\rm N_2}= 0.10$, $x_{\rm C_2H_6}= 0.55$ and $x_{\rm C_3H_8}= 0.05$. These data are then used in the algorithm
which assumes a sample composed only by the ternary mixture ($\rm CH_4$, $\rm N_2$, $\rm C_2H_6$). I found that, even with such a small amount
of propane, the derived abundances of $\rm N_2$ and the ratio $r_{46}$ are substantially affected: the error on  $x_{\rm N_2}$ is around
$20$\% (instead of $\sim 5$\% without $\rm C_3H_8$) while $r_{46}$ presents an error of about $100$\%. This numerical test emphasizes the sensibility
of the velocity of sound to composition. Obviously, to overcome this issue, the best solution is to introduce the measurement of another independent
physical quantity. 

%%%%%%%%%%%%%%%%%%%%%%%%%%%%%%%%%%%%%%%%%%%%%%%%%%%%%%%%%%%%%%%%%%%%%%%%%%%%%%%%%%%%%%%%%%%%%%%%%%%%%%%%%%%%%%%%%%%%%%%%%%%%%%%%%%%%
%%%%%%%%%%%%%%%%%%%%%%%%%%%%%%%%%%%%%%%%%%%%%%%%%%%%%%%%%%%%%%%%%%%%%%%%%%%%%%%%%%%%%%%%%%%%%%%%%%%%%%%%%%%%%%%%%%%%%%%%%%%%%%%%%%%

%%%%%%%%%%%%%%%%%%%%%%%%%%%%%%%%%%%%%%%%%%%%%%%%%%%%%%%%%%%%%%%%%%%%%%%%%%%%%%%%%%%%%%%%%%%%%%%%%%%%%%%%%%%%%%%%%%%%%%%%%%%%%%%%%%%%
\section{Conclusion}
\label{concl}

   In this work, I shown that the use of a realistic model, based on PC-SAFT, prevents the appearance of unphysical 
situations, like a speed of sound discontinuity, already
noticed in previous published papers \citep{hagermann_etal_2005,arvelo_lorenz_2013}. However, even such sophisticated models need to be constrained
by empirical data. Consequently, new laboratory measurements of sound speeds would be greatly useful, particularly in the case of mixtures.
The simultaneous measurements of refractive index, density, thermal conductivity, electromagnetic permittivity and speed of sound required by
Hagermann {\it et al}'s method is not easy to achieve and requires a complex set of sensors.\\
  Alternatively, the dynamic method proposed here, whereby the speed of sound is measured at different temperatures; provides a better
composition estimation with accuracy comparable to that of the measured
physical quantities (\textit{i.e.} a few percents), here the speed of sound. Indeed,
with only $50$ temperature measurements, I have shown that the errors in derived composition remain below
$10$\%. I also emphasized that the method employed
by \cite{hagermann_etal_2005} implicitly assumes the existence of a device implementing a temperature control by heating since they need 
thermal conductivity determinations. This fact suggest that, the two approaches could be combined in a future mission concept.

\section*{Acknowledgements}

I acknowledge Dr Ralph Lorenz for scientific discussions. I thank the anonymous Reviewer who 
improved the clarity of the paper with his/her remarks and comments. I also warmly thank my colleague Panayotis Lavvas 
for reading and improving my text.
 
%%%%%%%%%%%%%%%%%%%%%%%%%%%%%%%%%%%%%%%%%%%%%%%%%%



%\bibliographystyle{mnras}
%\bibliography{../BIB/bibliographie_planeto_2016} % if your bibtex file is called example.bib


%%%%%%%%%%%%%%%%%%%%%%%%%%%%%%%%%%%%%%%%%%%%%%%%%%

% Don't change these lines
\bsp	% typesetting comment
\label{lastpage}

%\end{linenumbers}

\end{document}